\documentclass[%
 aip,
 amsmath,amssymb,
 reprint,%
]{revtex4-1}

\usepackage{graphicx}
\usepackage{dcolumn}
\usepackage{bm}

\usepackage[utf8]{inputenc}
\usepackage[T1]{fontenc}
\usepackage{mathptmx}
\usepackage{etoolbox}
\usepackage[detect-all]{siunitx}
\usepackage[version=4]{mhchem}
\usepackage{fancyhdr}
\usepackage{hyperref}

\makeatletter
\def\@email#1#2{%
 \endgroup
 \patchcmd{\titleblock@produce}
  {\frontmatter@RRAPformat}
  {\frontmatter@RRAPformat{\produce@RRAP{*#1\href{mailto:#2}{#2}}}\frontmatter@RRAPformat}
  {}{}
}%
\makeatother
\begin{document}
\preprint{AIP/123-QED}

\title[Chip-scale sub-Doppler atomic spectroscopy enabled by a metasurface integrated photonic emitter]{Chip-scale sub-Doppler atomic spectroscopy enabled by a metasurface integrated photonic emitter}
\author{Alexander Yulaev}
\affiliation{Department of Chemistry and Biochemistry, University of Maryland, College Park, Maryland 20742, USA}
\affiliation{Physical Measurement Laboratory, National Institute of Standards and Technology, Gaithersburg, Maryland 20899, USA}
\author{Chad Ropp}%
\affiliation{Department of Chemistry and Biochemistry, University of Maryland, College Park, Maryland 20742, USA}
\affiliation{Physical Measurement Laboratory, National Institute of Standards and Technology, Gaithersburg, Maryland 20899, USA}
\author{John Kitching}
\affiliation{Time and Frequency Division, National Institute of Standards and Technology, 325 Broadway, Boulder, Colorado 80305, USA}
\author{Vladimir A. Aksyuk}
\affiliation{Physical Measurement Laboratory, National Institute of Standards and Technology, Gaithersburg, Maryland 20899, USA}
\author{Matthew T. Hummon}
\affiliation{Time and Frequency Division, National Institute of Standards and Technology, 325 Broadway, Boulder, Colorado 80305, USA}
 \email{matthew.hummon@nist.gov}

\date{\today}

\begin{abstract}
We demonstrate chip-scale sub-Doppler spectroscopy in an integrated and fiber-coupled photonic-metasurface device.
The device is a stack of three planar components: a photonic mode expanding grating emitter circuit with a monolithically integrated tilt compensating dielectric metasurface, a microfabricated atomic vapor cell and a mirror.  The metasurface photonic circuit efficiently emits a \SI{130}{\micro\meter}-wide ($1/e^2$ diameter) collimated surface-normal beam with only -6.3 dB loss and couples the reflected beam back into the connecting fiber, requiring no alignment between the stacked components. 
We develop a simple model based on light propagation through the photonic device to interpret the atomic spectroscopy signals and explain spectral features covering the full Rb hyperfine state manifold. 

\end{abstract}

\maketitle
\thispagestyle{fancy}
\fancyfoot[c]{\copyright 2024 Author(s). This article is distributed under a Creative Commons Attribution-NonCommercial 4.0 International (CC BY-NC) License. The following article has been submitted to Applied Physics Letters. After it is published, it will be found at \href{https://publishing.aip.org/resources/librarians/products/journals/}{Link}.}

Photonic integration enabling precision spectroscopy of atomic systems is a key component for next-generation quantum sensors with reduced size, weight, and power and improved manufacturability. \cite{kitching_chip-scale_2018, blumenthal_photonic_2020}  Sub-Doppler spectroscopy of warm atomic vapors is the basis for sensors for RF and DC electric fields,\cite{sedlacek_microwave_2012,jau_vapor-cell-based_2020} chip-scale optical atomic clocks,\cite{newman_architecture_2019} and realization of the meter. \cite{quinn_practical_2003} Much of the research of integrating warm atomic vapors with nanophotonics has focused on the interaction between atoms and the near-field evanescent tail of light guided in nanophotonic waveguides.\cite{ritter_atomic_2015,ritter_coupling_2016, stern_nanoscale_2013} While this results in a high level of integration and compact geometry, the small interaction volume in the proximity to the photonic component surfaces leads to substantial transit time broadening and large light shifts, limiting the resolution achievable for applications that require precision spectroscopy.  The development of suspended tapered waveguides permits larger optical modes in waveguide geometries, \cite{Zektzer2021} although the resolution of the spectroscopy is still limited to about \SI{100}{MHz} using thermal atomic vapors.

In general, sensors based on precision spectroscopy of atoms benefit from large free-space optical modes to maximize the number of atoms probed and reduce perturbations from interactions with nearby surfaces. This usually requires delivery of optical beams to positions of tens of micrometers to millimeters above the chip.  To date, the standard photonic element for achieving this is a grating emitter\cite{dakss_grating_1970}, which diffracts light from the chip into free space.  The grating emitters can be tailored for specific applications such as interrogation of trapped ions \cite{niffenegger_integrated_2020} or cooling beams for a magneto-optical trap. \cite{isichenko_photonic_2023,McGehee2021}  Recently, our group developed a photonic chip integrated with a microfabricated vapor cell that employed an extreme mode converting grating coupler \cite{kim_photonic_2018} to enable precision spectroscopy. \cite{hummon_photonic_2018}  The device generates a free-space optical mode with a diameter on the order of \SI{100}{\micro\meter} that enables precision sub-Doppler spectroscopy for laser frequency stabilization at the $10^{-11}$ level.  

One limitation of the device in Reference 16 is that the free-space beam emitted from the chip propagates a few degrees away from normal to the photonic chip, making alignment of the retro-reflecting mirror to generate a counter-propagating pump beam a complex step during assembly of the device.  An alternate geometry that uses a pair of grating couplers to generate overlapping counter-propagating beams is complicated by low overall coupling efficiency.\cite{bopp_nanophotonic_2019, hummon_dual_2021}  A grating coupler designed to emit the probe beam normal to the surface of the chip would simplify the assembly, allowing the photonic chip, vapor cell, and retro-reflector to be stacked directly on top of each other, enabling wafer level fabrication of fully integrated devices the same way chiplets are stacked in modern microelectronic systems on chip.  Current methods for generating surface-normal beams involve significant trade-offs in complexity, such as optically resonant gratings,\cite{yulaev_surface-normal_2023} multiple layers with precision alignment, \cite{michaels_inverse_2018,dai_highly_2015} or slanted etching.\cite{yin_observation_2020, wang_embedded_2005}

An alternative approach for generating a surface-normal beam is to co-integrate a grating emitter with a tilt correcting metasurface.  Metasurfaces use sub-wavelength features to precisely control degrees of freedom of light such as polarization state, phase, and intensity,\cite{chen_flat_2020} and their planar geometry allows them to be directly integrated on the surface of the chip. Metasurfaces, along with new simulation approaches such as inverse-design algorithms,\cite{molesky_inverse_2018} have opened new avenues for integrating photonic components with atomic systems.  Metasurfaces have been used to control the polarization and direction of propagation of light for generating magneto-optical traps.\cite{McGehee2021, jammi_alignment-free_2023} A high numerical aperture meta-lens was used for trapping and imaging single atoms.\cite{Hsu2022}  A metasurface has also been used with free-space optics for controlling the polarization of light in warm vapor spectroscopy.\cite{zhang_tailoring_2024} 
\begin{figure*}[htb!]
\includegraphics[width=17cm]{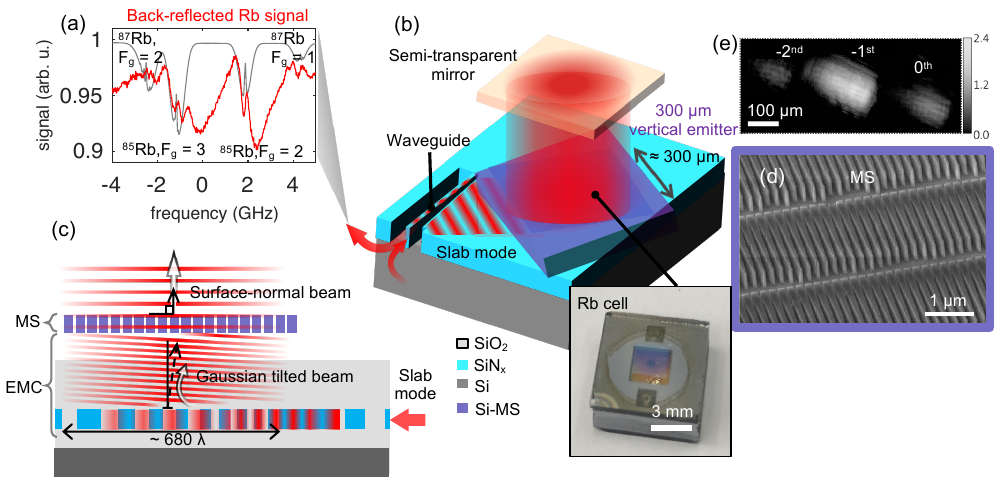}
\caption{\label{fig:one}The concept of sub-Doppler operation enabled using the metasurface photonic emitter. (a) Rb spectra at \SI{780}{nm}: the red curve depicts the back-reflected signal from the fiber-coupled photonic platform; the gray curve is a simultaneously measured reference Rb spectrum.  (b) The  3D schematic illustrating the MS-PIC emitter interrogating the planar microfabricated Rb cell. The single-mode waveguide is evanescently coupled to a slab mode allowing in-plane expansion of the photonic mode. The $(\SI{300}{\micro\meter})^2$ photonic grating emitter projects an $\approx\SI{130}{\micro\meter}$-wide collimated Gaussian beam out-of-plane followed by beam tilt correction to surface-normal using an MS optical wedge layer monolithically fabricated above the grating. The collimated beam passes through the Rb cell mounted directly on the photonic chip and is reflected from a mirror forming a pump-probe beam configuration. The inset depicts the optical image of the microfabricated Rb vapor cell.  (c) MS-PIC cross section schematic illustrating the formation of the surface-normal beam from a single-mode waveguide. (d) An SEM bird-eye micrograph of the MS array consisting of identical linearly tapered Si prisms forming an optical wedge metamaterial for beam tilting.  (e) A log-scale optical image of the 0th, -1st , and -2nd diffraction orders (DO) of the radiation projected into free space from the MS-PIC. The 0th and -2nd DO optical beam powers are $ \approx \SI{10}{\decibel}$ and $\approx\SI{13}{\decibel}$ below the surface-normal -1st order. }
\end{figure*}

Here we demonstrate a sub-Doppler atomic spectrometer that incorporates a tilt-compensating metasurface monolithically integrated with a waveguide-coupled, optimized photonic grating and a planar chip-scale microfabricated Rb vapor cell. First we describe the design and fabrication of the metasurface photonic integrated circuit (MS-PIC) surface emitter.  We develop a simple model to describe the light propagation through the device and use it to analyze the resulting Rb spectra.

Figure \ref{fig:one} shows an overview of the integrated MS-PIC assembly and details of its components.   The device begins with probe light at \SI{780}{nm} propagating in a single-mode silicon nitride waveguide, to which it is coupled from a single-mode optical fiber through an inverse taper edge coupler.  The photonic mode is expanded laterally through evanescent coupling from the single-mode waveguide to a \SI{100}{\micro\meter}-wide collimated 1D Gaussian slab mode. The slab mode impinges on a large apodised free-space emitter projecting a collimated 2D Gaussian beam at $\approx\SI{11.2}{\degree}$ in glass relative to the chip’s normal (Fig. 1(b,c)). The details of photonic mode expansion design due to evanescent coupling and grating beam projection can be found elsewhere. \cite{kim_photonic_2018,ropp_scalable_2023} 
The metasurface (Fig. 1(c,d)) is situated directly above the grating on top of a \SI{3}{\micro\meter}-thick \ce{SiO2} cladding and is designed to uniformly refract the tilted beam to surface normal. Each of the metasurface’s identical unit cells (Fig. 1(d)) contains a long and narrow Si prism with a linearly varying width and a length of \SI{3.1}{\micro\meter} along the beam tilt direction such that the -1st diffraction order is oriented vertically. The sub-wavelength widths and spacings between the prisms are designed to form an optical metamaterial imparting a linearly varying phase to maximize the light intensity of the vertical diffraction order akin to a blazed grating (Fig. 1(e)). 


We fabricated the MS-PIC emitter in two separate steps.  First, we fabricated the photonic emitter and experimentally measured the polar and azimuthal angles of the projected free-space beam by analyzing the light beam profiles collected at different heights above the chip.  Second, we designed the tilt-correcting MS based on the experimentally quantified polar angle and fabricated it directly on the top \ce{SiO2} cladding above the photonic emitter. Finally, we quantified the coupling efficiency of the whole PIC emitter and power distribution across the diffraction orders.
Figure 1(e) shows the intensity profiles of the -2nd, -1st, and 0th diffraction orders in a logarithmic scale at an imaging plane a few millimeters above the MS-PIC surface. The brightest spot is the -1st diffraction order that is oriented vertically and used for the Rb sub-Doppler spectroscopy. The propagation direction, intensity profile, and the power of the -1st diffraction order ($\approx\SI{-6.2}{\decibel}$ relative to the power in an input fiber) agrees well with the simulated result (\SI{-2.4}{\decibel} from the meta-surface and $\approx\SI{-2}{\decibel}$ from the grating coupler and the fiber edge coupler, each). The amount of power in each of the 0th and -2nd order modes is suppressed by at least a factor of 10 relative to the power in the -1st order beam.  

\begin{figure*}[htb!]
\includegraphics[width=17cm]{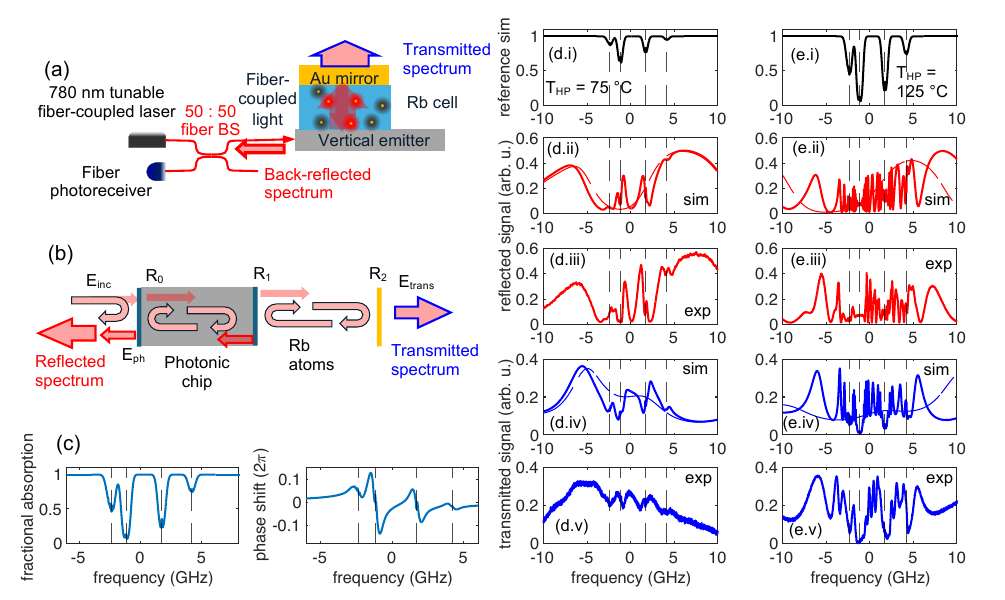}
\caption{\label{fig:two}Photonic spectroscopy system. (a)  Schematic of optical setup for spectroscopy.  (b) Dual-cavity system for modeling spectroscopic features. (c) Simulated absorption and dispersive atomic responses. (d,e)  Comparison of simulated and measured spectra at low and high atomic density. Panel (i) show bare simulated atomic absorption. Panels (ii), (iv) show the simulated reflection and transmitted signal in solid lines.  The dashed line shows the simulated bare (empty) cavity + MS-PIC response.  Panels (iii), (v) show the corresponding measured data.}
\end{figure*}

To complete the MS-PIC hybrid spectrometer we place a microfabricated rubidium vapor cell on top of the MS-PIC surface. The vapor cell is \SI{3}{mm} in height and contains rubidium inside of an evacuated chamber.\cite{kitching_chip-scale_2018}  A $\approx\SI{90}{\percent}$ reflective gold mirror placed directly on top of the Rb vapor cell retro-reflects the probe beam back into the MS-PIC emitter and waveguide.   With the vapor cell and reflector placed on top of the meta-surface photonic emitter, we measure a total round trip transmission efficiency of $\approx\SI{-16}{\decibel}$, with contributions of about \SI{-13}{\decibel} from the MS-PIC and about \SI{-3}{\decibel} from mode mismatch of the retroreflected beam. 



Figure 2(a) shows the optical setup we use to collect spectra from the fully assembled device. We use a simple 50:50 fiber beam splitter to connect the fiber-coupled laser and photoreceiver to the PIC.  The PIC and Rb vapor cell sit on a heated baseplate that maintains a device temperature in the range of \SIrange[]{70}{125}{\celsius} to allow for tuning of the Rb vapor density within the cell.  We record both the retro-reflected fiber coupled signal as well as an auxiliary transmission measurement detected with a free-space photodiode placed above the vapor cell. Figure 1(a) shows a typical set of spectra, taken at a baseplate operating temperature of \SI{75}{\celsius}.  The reflected signal clearly shows a dispersive response near the atomic absorption features, indicating the signal is dominated by the phase shift the atoms impart on the light.  This is a common feature in experiments where the signal results from the interference of light from two paths that experience differential atomic phase shifts, such as atomic Mach Zehnder interferometers,\cite{ritter_atomic_2015} atoms in optical resonators,\cite{papoyan_straightforward_2017,ritter_coupling_2016} or atomic diffractive elements.\cite{Stern2019}    In our case, we model our system as a pair of optical cavities, shown schematically in Fig. 2(b).  The main optical cavity consists of the Rb vapor cell and is formed from reflections from the surface of the photonic chip ($R_1 \approx 0.1$) and the partial reflector placed on top of the vapor cell ($R_2\approx 0.9$). It has an effective optical path length of about \SI{5}{mm}.  The second, weaker cavity is formed from reflections between the input coupling facet on the photonic chip ($R_0 \approx 0.04$) and the grating output coupler, and has an optical path length of about \SI{11.8}{mm}.  The parameters of this second weaker cavity are estimated from reflection measurements of the photonic chip made with the Rb vapor cell removed.  We adopt the circulating field formalism \cite{siegman_resonance_1986} to calculate the transmitted and reflected beams from the atom-dual-cavity system.  The absorption and phase shift due to the atoms is modeled using the atomic susceptibility.  To calculate the absorption spectrum, we use a saturated absorption lineshape model,\cite{pappas_saturation_1980, wang_observation_2015} and the phase shift is then calculated from the atomic absorption using the modified Kramers-Kronig relation to take into account saturation and hyperfine optical pumping effects. \cite{bolton_modification_1969, troup_use_1973, ryan_density_1986}  Figure \ref{fig:two}(c) shows typical absorption and dispersion profiles used for the device simulations.   In general, the electric field returned from the photonic chip $E_\text{ph}$ is more complex than the field calculated using the simple two-cavity model presented here due to additional reflections present on the photonic chip, possibly from fabrication imperfections in the waveguide structure. In the data presented in Fig. {\ref{fig:two}(d,e)}, we find the best agreement between our model and experiment when we include a field from one additional on-chip reflection.

We analyze the system for two cases, corresponding to low (\SI{3e11}{cm^{-3}}) or high (\SI{5e12}{cm^{-3}}) atomic density in the vapor cell. The results of the simulation are shown in Fig. 2(d). Panel (d.i) corresponds to the simulated bare atomic absorption profile for each operating temperature.  The centers of the four Rb absorption peaks are each indicated with black dashed vertical lines.  Panel (d.ii)/(d.iv) shows the simulated reflected (transmitted) signal in a solid line.  The dashed line corresponds to the reflected (transmitted) signal due to the dual cavity system in absence of the contribution from the atomic signal.  We see that at low atomic density, the signal is dominated by atomic dispersion near the atomic resonances and far from resonance (greater than \SI{1}{GHz}) the response returns to the bare dual cavity system.  We observe good qualitative agreement between the simulated and measured (panels d.iii, d.v) spectra for both the reflected and transmitted spectra.   Figure 2(e) shows the same measurements for high atomic density.  At high atomic density, the strong atomic absorption dominates the spectral response near the four atomic resonances.  Away from resonance we observed oscillations of the signal due to the fall off of the strong atomic dispersive response.  At an operating temperature of \SI{125}{\celsius}, we observe the dispersive contributions for detunings larger than \SI{5}{GHz}.
\begin{figure}[tb!]
\includegraphics[width=8.5cm]{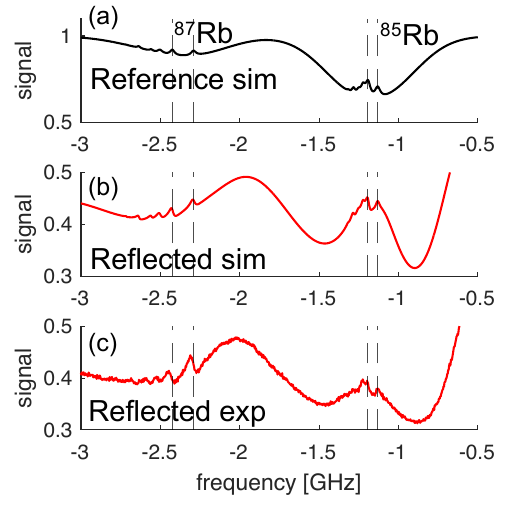}
\caption{\label{fig:three} Sub-Doppler spectroscopy of Rb. Panel (a) shows the simulated absorption spectrum of a free-space reference beam.  Panels (b) and (c) show the reflected spectrum for the MS-PIC device for both the simulation and experiment, respectively.  The dispersive sub-Doppler features are clearly resolved in the \ce{^{87}Rb} transition, with peak-to-peak linewidths of approximately \SI{40}{MHz}.  The vertical dashed lines show positions of selected sub-Doppler absorption features. } 
\end{figure}

For many applications, spectroscopic features much narrower than the Doppler-broadened linewidth are advantageous.  Figure 3 shows a spectrum where the excited state hyperfine levels and cross-over resonances in  \ce{^{87}Rb} are well resolved.  The sub-Doppler features are generated from velocity selective optical pumping effects from a pair of counterpropagating beams inside the vapor cell.  The dispersive lineshapes have a peak-to-peak linewidth of approximately \SI{40}{MHz}, likely due to power broadening from the probe laser beam intensity.  Using the same two-cavity model, we observe good qualitative agreement between the simulated spectrum and the observed experimental reflected spectrum, shown in Fig. \ref{fig:three}(b) and (c), respectively.  We note that the sub-Doppler features in the \ce{^{87}Rb} transition have a more dispersive character, while the corresponding lines in the \ce{^{85}Rb} transition have more absorptive character.  This is due to the position of the atomic resonance relative to the overall cavity transmission.  Here the \ce{^{87}Rb} line falls on the side of a cavity transmission, while the \ce{^{85}Rb} corresponds to the peak of the cavity resonance.  

In some cases, the dispersive feature can be advantageous, allowing for laser stabilization to the center of the feature without the need for additional frequency modulation to generate an error signal.  However, if the cavity resonance shifts relative to the atomic resonance, the lock point could become unstable. To eliminate this dependence of the lock point on the position of the cavity resonance, it would be preferable to minimize the reflections and interference effects that contribute to the dispersive features.  One possible route would be to use 2-d polarization-dependent gratings that couple light of orthogonal polarizations into separate waveguides.\cite{watanabe_2-d_2019,sebbag_demonstration_2021}  This type of polarization dependent grating in conjunction with a metasurface optic that rotates the polarization by $\lambda/4$ on a single pass would direct the probe beam into a waveguide separate from the pump beam after it has interrogated the atoms in the vapor cell, similar to the common free-space optic sub-Doppler absorption setups.\cite{liang_compact_2015,cheng_novel_2023}  Using the metasurface and grating emitter integration methods described here, fabricating such a device to enable a separate waveguide channel for readout of the atomic absorption looks to be a promising path for suppressing inteference effects in chip-scale integrated sub-Doppler spectroscopy.

In conclusion, by integrating a tilt compensating metasurface with a photonic mode expanding grating emitter we have performed sub-Doppler spectroscopy of a warm Rb vapor using an alignment-free assembly of the MS-PIC, microfabricated Rb vapor cell, and retroreflecting mirror.  In addition, we have developed a simple two-cavity model to interpret the observed atomic spectra for a range of atomic densities spanning a factor of ten. Future co-integration of on-chip lasers\cite{xiang_high-performance_2021} and detectors\cite{lin_monolithically_2022} will enable manufacturable quantum sensors based on precision spectroscopy of warm atomic vapors.






\begin{acknowledgments}
Dr. Alexander Yulaev acknowledges support under the Professional Research Experience Program (PREP), funded by the National Institute of Standards and Technology and administered through the Department of Chemistry and Biochemistry, University of Maryland. Authors also thank Dr. Yang Li and Dr. David Carlson for reading the manuscript and making insightful comments. Research performed in part at the NIST Center for Nanoscale Science and Technology. This work was partially supported by funding from the NIST-on-a-Chip program.
\end{acknowledgments}

\section*{Data Availability Statement}

The data that support the findings of this study are available from the corresponding authors upon reasonable request.
\bibliography{references}

\end{document}